\documentclass{article}
\pdfoutput=1
\usepackage[utf8]{inputenc}
\usepackage[margin=1in]{geometry}
\usepackage{amssymb,amsfonts,amsmath,comment,mleftright,enumerate,graphicx,hyperref,xcolor,numprint}
\usepackage[makeroom]{cancel}
\usepackage{multicol}

\newtheorem{definition}{Definition}

\newcommand{\paren}[1]{\left(#1\right)}
\newcommand{\bracket}[1]{\left[#1\right]}
\renewcommand{\brace}[1]{\left\{#1\right\}}
\newcommand{\ang}[1]{\left\langle#1\right\rangle}

\newcommand{\Z}{\mathbb{Z}}
\newcommand{\ZM}[1]{\Z/#1\Z}
\newcommand{\F}{\mathbb{F}}

\newcommand{\M}[1]{\begin{bmatrix}#1\end{bmatrix}}
\newcommand{\SM}[1]{\bracket{\begin{smallmatrix}#1\end{smallmatrix}}}

\renewcommand{\min}[1]{\mathrm{min}\paren{#1}}
\renewcommand{\max}[1]{\mathrm{max}\paren{#1}}

\newcommand{\false}{\mathtt{FALSE}}

\newcommand{\cyc}{\bigtriangleup}
\newcommand{\tp}{\intercal}
\newcommand{\sw}[1]{\phi_{#1}}
\newcommand{\id}{\mathrm{id}}
\newcommand{\mmt}[1]{\mathcal{T}^{\langle #1 \rangle}}
\newcommand{\lexlt}{<_{\mathrm{lex}}}
\newcommand{\listjoin}{\parallel}
\newcommand{\trace}[1]{\mathrm{tr}\paren{#1}}
\newcommand{\degroote}[1]{\boldsymbol{\Gamma}_{#1}}

\title{Ruling Out Low-rank Matrix Multiplication Tensor Decompositions with Symmetries via SAT}
\author{Jason Yang}
\date{}

\begin{document}

\maketitle

\begin{abstract}
We analyze rank decompositions of the $3\times 3$ matrix multiplication tensor over $\ZM{2}$. We restrict our attention to decompositions of rank $\le 21$, as only those decompositions will yield an asymptotically faster algorithm for matrix multiplication than Strassen's algorithm.
To reduce search space, we also require decompositions to have certain symmetries.
Using Boolean SAT solvers, we show that under certain symmetries, such decompositions do not exist.
\end{abstract}

\begin{multicols}{2}

\section{Introduction}
Matrix multiplication is a fundamental operation of linear algebra, as well as the primary performance bottleneck in numerous graph theory algorithms and machine learning models.
In 1969, Strassen discovered the first algorithm for $N\times N$ matrix multiplication that ran faster than $O(N^3)$ time, with time complexity $O(N^{2.808})$ \cite{strassen}. Since then, an entire subfield of computer science has been created around this problem, with a long list of improvements \cite{survey}; currently the fastest algorithm has time complexity $O(N^{2.371552})$ \cite{record}.
However, nearly all algorithms after Strassen have such enormous overhead that they are too slow in practice and have thus never been implemented.

All asymptotically fast algorithms discovered so far rely on a divide-and-conquer scheme, in which one splits the input matrices $X$ and $Y$ into blocks, recursively computes products of linear combinations of these blocks, then constructs blocks of the target matrix $Z=XY$ as linear combinations of those products.
WLOG all algorithms in the arithmetic circuit model obey this form; in particular, using divisions does not help \cite{division}.

In the most general kind of scheme, the left input $X$ is split into $n\times k$ many blocks, and the right input $Y$ is split into $k\times m$ many blocks. It can be shown \cite{survey} that finding coefficients for the linear combinations involved in the divide-and-conquer such that the resulting matrix $Z$ is correct is equivalent to finding a list of matrix triplets $D=[(A^{(r)}\in\F^{n\times k},B^{(r)}\in\F^{k\times m},C^{(r)}\in\F^{m\times n})]_{0\le r<R}$ s.t.
\begin{equation}
    \mmt{n,k,m}_{a,b,c,d,e,f}=\sum_{r=0}^{R-1} A^{(r)}_{a,b}B^{(r)}_{c,d}C^{(r)}_{e,f}\ \forall a,b,c,d,e,f,
\end{equation}

where $\F$ denotes the field we are working over, $R$ is the number of matrix triplets in $D$, and $\mmt{n,k,m}$ is a $n\times k\times k\times m\times m\times n$ tensor (multidimensional array) known as the $(n,k,m)$-\textit{matrix multiplication tensor} and is defined as
\begin{equation}
    \mmt{n,k,m}_{i,j,j,k,k,i}=1\ \forall i,j,k \textrm{; all other elements are } 0.
\end{equation}

The list of matrix triplets $D$ is known as a \textit{decomposition} of $\mmt{n,k,m}$, and the number of matrix triplets $R$ in $D$ is called the \textit{rank of the decomposition}.
The smallest $R$ such that a rank-$R$ decomposition of $\mmt{n,k,m}$ exists is known as the \textit{rank of the tensor} $\mmt{n,k,m}$.

The equation involving $\mmt{n,k,m}, A, B, C$ can also be expressed as
\begin{equation}
    \mmt{n,k,m}=\sum_{r=0}^{R-1} A^{(r)}\times B^{(r)}\times C^{(r)},
\end{equation}
where $\times$ denotes the outer product (equivalent to $\texttt{numpy.multiply.outer()}$ from NumPy).

A rank-$R$ decomposition of $\mmt{n,k,m}$ yields a $O(N^{3\log_{nkm} R})$-time algorithm for multiplying two $N\times N$ matrices \cite{survey}. 
Strassen's algorithm corresponds to a rank-7 decomposition of $\mmt{2,2,2}$, and it has running time $O(N^{\log_2 7})\approx O(N^{2.808})$ \cite{strassen}.
Much effort since 1969 has been devoted to finding low-rank decompositions of tensors $\mmt{n,k,m}$ that improve Strassen's algorithm.
However, rank 7 is optimal for $\mmt{2,2,2}$ \cite{winograd}, and analyzing other matrix multiplication tensors proved to be very difficult, with no improvements found until Pan in 1978 discovered a rank-143640 decomposition of $\mmt{70,70,70}$ \cite{pan78}.

Every record-breaking algorithm since Pan, including the current record, has relied on \textit{border rank decompositions}, which do not evaluate to the target tensor exactly but get arbitrarily close to it, from which exact decompositions can be constructed. Over the decades, additional techniques have allowed better border rank decompositions to be constructed via increasingly indirect means \cite{survey}, leading to algorithms with much better asymptotic complexity than algorithms derived from any known explicitly constructed (non-border rank) decomposition.

However, these algorithms only obtain good asymptotics for tensors $\mmt{n,k,m}$ with enormously large $n$, $m$, and $k$, which creates prohibitively high overhead.
For this reason, there is still interest in finding explicit decompositions of small $\mmt{n,k,m}$, with significant progress in recent years due to the help of computers \cite{smirnov} \cite{alphatensor} \cite{flip}.

We are interested in the tensor $\mmt{3,3,3}$, as it is the smallest tensor (corresponding to square matrix multiplication) whose exact rank is still uncertain. Currently, its rank is known to be between 19 \cite{333lower} and 23 \cite{laderman}.
To reduce search space, we restrict decompositions to be over the integers modulo 2, as well as artificially force decompositions to satisfy certain symmetry constraints, the motivation of which will be given later.

This work is our third paper on this subject. In the previous two works \cite{part1} \cite{part2}, we searched for rank-$\le 23$ decompositions of $\mmt{3,3,3}$ over the integers mod 2 under certain symmetry constraints, using a hand-crafted meet-in-the-middle search.
In this work, we use a preexisting SAT solver and restrict our attention to rank-$\le 21$ decompositions, as 21 is the highest possible rank for a decomposition of $\mmt{3,3,3}$ to yield an asymptotically faster algorithm than Strassen's algorithm ($\log_3 21 < \log_2 7 < \log_3 22$).

\subsection{Related work}
Our main inspiration to impose symmetry constraints on decompositions is from \cite{ballard}, which enforced cyclic symmetry (to be defined later), then used nonlinear optimization followed by sparsification to find several rank-23 decompositions over the integers (not mod anything) of $\mmt{3,3,3}$.

Additionally, numerous previous works have searched for decompositions of $\mmt{3,3,3}$ over the integers mod 2, using approaches such as SAT \cite{courtois} \cite{oh}, machine learning \cite{alphatensor}, and randomized local search \cite{heule} \cite{flip} \cite{adaflip}, with some new low-rank decompositions being found.
In contrast to these approaches, we focus on exhaustive analysis and avoid any form of randomization that might affect correctness.

It is also worth mentioning that there have been other approaches that do not restrict decompositions to mod 2, such as alternating least-squares over floating-point values \cite{smirnov} and hand-crafted decompositions over the integers \cite{hopcroft-kerr} \cite{sykora} \cite{pan78}. Since these approaches are difficult to generalize and are also not exhaustive, we do not pursue them here.

\subsection{Symmetries}
Matrix multiplication tensors are known to allow an arbitrary decomposition of them to be transformed into many other valid decompositions.
Given a decomposition $\mmt{n,n,n}=\sum_r A^{(r)}\times B^{(r)}\times C^{(r)}$ corresponding to \textit{square} matrix multiplication, it can be verified by hand that performing the following transformations on each matrix triplet yields another valid decomposition of $\mmt{n,n,n}$:

\begin{itemize}
    \item cyclic: $\cyc((A,B,C))=(B,C,A)$
    \item transpose: $\tp((A,B,C))=(C^\top,B^\top,A^\top)$
    \item sandwich: $\sw{U,V,W}((A,B,C)) \\ =(UAV^{-1},VBW^{-1},WCU^{-1})$
    
    for invertible $U,V,W\in \F^{n\times n}$
\end{itemize}

The group $\degroote{n}$ generated by these functions is known as the set of De Groote symmetries \cite{degroote} and can be derived via algebraic properties of matrix trace\footnote{
    For arbitrary $n\times n$ matrices $X,Y,Z$, we have $\sum_{a,b,c,d,e,f} \mmt{n,n,n}_{a,b,c,d,e,f}X_{a,b}Y_{c,d}Z_{e,f}=\trace{XYZ}$; then $\trace{XYZ}
    =\trace{YZX}
    =\trace{Z^\intercal Y^\intercal X^\intercal}
    =\trace{(UXV^{-1})(VYW^{-1})(WZU^{-1})}$.
}.
The reason we focus on square matrix multiplication is that non-square $\mmt{n,k,m}$ do not have cyclic and transpose symmetry (although they still have sandwich symmetry).

If performing a transformation in $\degroote{n}$ on a decompostion yields the same decomposition up to permutation of the matrix triplets, we say that the \textit{decomposition is symmetric} under that transformation.

Our main motivation for enforcing decompositions to be symmetric under one or more transformations in $\degroote{n}$, besides reducing search space, is that many low-rank decompositions of small $\mmt{n,n,n}$ happen to be symmetric under some subgroup of $\degroote{n}$, such as those of Strassen\footnote{
    The rank-7 decomposition of $\mmt{2,2,2}$ that Strassen's algorithm corresponds to is symmetric under $\ang{\cyc,\phi_{F,F,F}}$ with $F=\M{0&1\\1&0}$.
} \cite{strassen}, Ballard et. al. \cite{ballard}, and others \cite{heule}.
Because of this, one might suspect that a low-rank decomposition of $\mmt{3,3,3}$ might also have such a symmetry. 

\subsection{Previous Work}
In our past two papers \cite{part1} \cite{part2}, we used an approach where given a specific symmetry group, we first enumerate all orbits of triplets of $3\times 3$ matrices, encode their resulting tensors as binary strings (while keeping track of their rank), then use meet-in-the-middle to find some subset of these binary strings that XOR to a target while having total rank below 23.

We analyzed symmetry groups of the form $\ang{f}$, $\ang{\cyc,f}$ and $\ang{\cyc,\tp,f}$ for an arbitrary transformation $f\in \degroote{3}$. Within each of these sets of groups, we ignored conjugates, since if a decomposition $D:=[t_i\in (\F^{n\times n})^3]_i$ is invariant under a subgroup $G\subseteq\degroote{n}$, than $D':=[f(t_i)]_i$ is invariant under $fGf^{-1}$.

We found a few rank-23 decompositions over $\ang{\sw{\SM{1&0&1\\1&1&0\\0&0&1},I,I}\circ\cyc}$,
$\ang{\cyc,\sw{\SM{1&0&1\\1&1&0\\0&0&1},\SM{1&0&1\\1&1&0\\0&0&1},\SM{1&0&1\\1&1&0\\0&0&1}}}$,
$\ang{\cyc,\sw{\SM{0&1&0\\1&0&0\\0&0&1},I,I}\cdot\tp}$, and
$\ang{\cyc,\tp,\sw{\SM{0&1&0\\1&0&0\\0&0&1},\SM{0&1&0\\1&0&0\\0&0&1},I}}$\footnote{The first two groups mentioned in this list happen to be conjugate to $\ang{\cyc,\sw{\SM{0&0&1\\1&0&1\\0&1&1},\SM{0&0&1\\1&0&1\\0&1&1},\SM{0&0&1\\1&0&1\\0&1&1}}}$, the mod 2 reduction of the first group that \cite{ballard} presented in their paper.

Additionally, the first two groups in this list are conjugate to each other, and the last two groups are conjugate to each other (but the first group is not conjugate with the third).
The reason none of these four groups were ignored was because we did not account for conjugacy between groups of different types; the first and second groups mentioned in this list are of type $\ang{f}$ and $\ang{\cyc,f}$, respectively; the third and fourth are of type $\ang{\cyc,f}$ and $\ang{\cyc,\tp,f}$, respectively.},
but proved that none of these groups yield rank below 23.

We also found that no subgroup of the form $\ang{\cyc,\tp,f}$ has any rank-23 decompositions, except possibly $\ang{\cyc,\tp}$. This means that no group that is a strict superset of $\ang{\cyc,\tp}$ has any rank-23 decompositions. Additionally, when focusing more narrowly on rank-21 decompositions, the only possible subgroups of the form $\ang{\cyc,f}$ were $\ang{\cyc}$, $\ang{\cyc,\tp}$, and $\ang{\cyc,\sw{\SM{1&1&0\\0&1&0\\0&0&1},\SM{1&1&0\\0&1&0\\0&0&1},\SM{1&1&0\\0&1&0\\0&0&1}}}$\footnote{\label{part2-table}
    The full tables of our previous search results are available at the appendix of \url{https://github.com/coolcomputery/Matrix-Multiplication-Tensor-Decomposition/blob/4508649a56a2861fd3a262c1159feba959d48d60/full-part2-report.pdf}.
    If a row has its $R$ field equal to or greater than 21 and not N/A, that means that the symmetry group corresponding to that row does not have a rank-$\le 21$ decomposition of $\mmt{3,3,3}$.
}. Previously we were only able to rule out decompositions of rank 7, 11, and 14 for these three groups, respectively.
We analyze these subgroups further in this paper.

For the form $\ang{f}$, there are many more such groups that might have rank-21 decompositions. We leave these groups to future work.

\subsection{Results}
\label{results}
As stated in the previous subsection, we investigate the symmetry groups $\ang{\cyc}$, $\ang{\cyc,\tp}$, and $\ang{\cyc,\sw{F,F,F}}$ with $F=\M{1&1&0\\0&1&0\\0&0&1}$.

By using the Z3 SAT solver,
we found that over $\ZM{2}$, there are no $\ang{\cyc,\tp}$ or $\ang{\cyc,\sw{F,F,F}}$ -symmetric decompositions of $\mmt{3,3,3}$ with rank $\le 21$.
This means there are no decompositions over $\ZM{2}$ that are symmetric under $\ang{\cyc,\tp}$ or $\ang{\cyc,\sw{F,F,F}}$ which yield an asymptotically faster algorithm than Strassen's algorithm.
Combining this with our previous work, we have that any $\ang{\cyc}$-symmetric rank-$\le 21$ decomposition cannot have any other nontrivial symmetries.

We also tried searching the less restrictive $\ang{\cyc}$-symmetric decompositions of $\mmt{3,3,3}$ in the hope of obtaining a stronger result than the current general lower bound of rank 19 \cite{333lower}, but we were only able to rule out rank $\le 15$. We tried running $\ang{\cyc}$-symmetry on rank 16 but did not get any partial progress after running for more than $\numprint{900 000}$ seconds.

Table \ref{tab:results} summarizes our results.
Our code is available at \begin{center}\url{https://github.com/coolcomputery/Matrix-Multiplication-Tensor-Decomposition}.\end{center}


\section{Notation}
\begin{itemize}
    \item $\listjoin$ denotes list concatenation.
    \item For an $n$-element list $L=\bracket{L_0,\dots L_{n-1}}$, $L_{i:j}$ denotes the slice $\bracket{L_i,\dots L_j}$, and $L_{i:}$ is short for $L_{i:n-1}$.
\end{itemize}

\section{SAT Formulation}

Our initial attempt to enforce symmetry constraints was to first encode general tensor decomposition mod 2 (using $\land$ as multiplication and XOR as addition) as SAT, then constrain variables to be equal to auxiliary variables so that the decomposition was forced to obey a given symmetry, e.g. $abc=1 \land (a=d \land b=d \land c=d)$. We later found that directly inputting those auxiliary variables into the decompostion (which we call ``inlining"), e.g. replacing the previous expression with $ddd=1$, gave a $\sim 1.5-2$ times speedup on difficult instances and a $\sim 10$ times speedup on easy instances.

In all SAT instances for searching a rank-$R$ decomposition, we enforce each matrix triplet $(A,B,C)$ in a decomposition to not be all-zero, as otherwise $A\times B\times C$ would be all-zero and removing this matrix triplet would result in a decomposition with rank $R-1$. We employ additional techniques to avoid ``rank redundancy".

We also account for permutation invariance between matrix triplets: $\sum_r A^{(r)}\times B^{(r)}\times C^{(r)}=\sum_r A^{(\sigma(r))}\times B^{(\sigma(r))}\times C^{(\sigma(r))}\ \forall \sigma\in\mathrm{Sym}(R)$. We do this by enforcing \textit{lexicographic} (abbr. ``lex") inequalities between adjacent matrix triplets. To compare matrices or tensor-like objects (e.g., matrix triplets\footnote{To compare matrix triplets $(A,B,C)$, we flatten the 3D array $[A,B,C]$ to $\vec{A}\listjoin\vec{B}\listjoin\vec{C}$, where $\vec{M}$ denotes the flattening of matrix $M$.}), we flatten them in row-major order, equivalent to $\texttt{numpy.flatten()}$, and lexicographically compare the resulting lists.

Since we are working over mod 2, we can prevent duplicate matrix triplets by making all lex inequalities strict.
The lex less-than inequality is defined recursively as follows:
\begin{align*}
([] \lexlt []):= & \false
\\
(a \lexlt b):= & (\neg a_0 \land b_0) \\ & \lor \paren{a_0=b_0 \ \land \ (a_{1:}\lexlt b_{1:})}
\end{align*}

We also do not have to address scaling symmetry $(A,B,C)\mapsto(\alpha A,\beta B,\gamma C)$ for $\alpha\beta\gamma=1$, as over mod 2 $(\alpha,\beta,\gamma)$ is forced to be $(1,1,1)$, so the previously mentioned strict lex inequalities already address duplicate matrix triplets.


\subsection{Cyclic symmetry}
For any given symmetry group, we can enumerate all possible orbits of a matrix triplet under the group by fixing the stabilizer of the triplet.

\begin{definition}
Under $\ang{\cyc}$ symmetry, we have the following orbits:
\begin{itemize}
    \item $\Theta_\id(A,B,C)=[(A,B,C),(B,C,A),(C,A,B)]$
    \item $\Theta_\cyc(D)=[(D,D,D)]$
\end{itemize}

$R_\id$ denotes the number of $\Theta_\id$-orbits and $R_\cyc$ the number of $\Theta_\cyc$-orbits in a $\ang{\cyc}$-symmetric decomposition.

Our resulting decomposition is
\begin{align*}
& \paren{\listjoin_{0\le r<R_\cyc} \Theta_\cyc(D^{(r)})} \\
\listjoin & \paren{\listjoin_{0\le r<R_\id} \Theta_\id(A^{(r)},B^{(r)},C^{(r)})},
\end{align*}
with total rank $R=3R_\id+R_\cyc$.
\end{definition}

To reduce rank redundancy, we notice that if we have two $\Theta_3$-orbits of the form $\Theta_3((A,B,C))$ and $\Theta_3((A,B,C'))$, we can merge them into $\Theta_3((A,B,C+C'))$, since 
\begin{align*}
    & (A\times B\times C+B\times C\times A+C\times A\times B) \\
    & +(A\times B\times C'+B\times C'\times A+C'\times A\times B)\\ 
    = &
    A\times B\times (C+C') \\
    & +B\times (C+C')\times A \\
    & +(C'+C)\times A\times B;
\end{align*} this means we can force all $(A^{(r)},B^{(r)})$ to be distinct, and thus force them to be lex strictly increasing w.r.t $r$, which is stricter than having all $(A^{(r)},B^{(r)},C^{(r)})$ be lex strictly increasing.

By incorporating this observation, we can force an arbitrary $\ang{\cyc}$-symmetric decomposition into a canonical form as follows:
\begin{enumerate}
    \item Replace each $(A^{(r)},B^{(r)},C^{(r)})$ with the lex minimum triplet in its orbit $\Theta_\id(A^{(r)},B^{(r)},C^{(r)})$;
    \item Merge triplets $(A^{(r)},B^{(r)},C^{(r)})$ by the $(A^{(r)},B^{(r)})$ part (e.g., using a Python dictionary), adding together the $C^{(r)}$ parts;
    \item Sort $D^{(r)}$ in strictly increasing lex order w.r.t. $r$
    \begin{itemize}
        \item[$\circ$] We can force strict lex inequalities because duplicate $D^{(r)}$ will yield duplicate orbits that cancel each other out over $\ZM{2}$.
    \end{itemize}
    \item Sort $(A^{(r)},B^{(r)},C^{(r)})$ in strictly increasing lex order.
\end{enumerate}

The resulting decomposition has rank at most that of the original (possibly less), so WLOG we can force all decompositions we search to satisfy this canonical form.

To enforce this form, we can require each orbit representative $(A,B,C)$ to be the \textit{strict} lex minimum triplet in $\Theta_\id(A,B,C)$, since if two matrix triplets in that orbit were equal, $\Theta_\id(A,B,C)$ would degenerate into 3 copies of a $\Theta_\cyc$-orbit.
To account for both rank redundancy and lex ordering of $(A^{(r)},B^{(r)},C^{(r)})$ at the same time, we can force all $(A^{(r)},B^{(r)})$ to be lex strictly increasing w.r.t. $r$, which is more strict than $(A^{(r)},B^{(r)},C^{(r)})$ to be lex strictly increasing.

Thus, we add the following set of constraints to the tensor decomposition equation for our SAT (over all applicable indices $r$):
\begin{itemize}
    \item $(A^{(r)},B^{(r)},C^{(r)})\lexlt t$ 
    
    $\forall t\in \big\{
    (B^{(r)},C^{(r)},A^{(r)}),
    (C^{(r)},A^{(r)},B^{(r)})\big\}$


    
    \item $(A^{(r)},B^{(r)})\lexlt (A^{(r+1)},B^{(r+1)})
    $
    
    \item $D^{(r)}\lexlt D^{(r+1)}
    $
\end{itemize}

When searching for all $\ang{\cyc}$-symmetric decompositions of rank $\le R$, we test over all combinations of orbit counts $(R_\id,R_\cyc)$ s.t. total rank is $\le R$, i.e., $3R_\id+R_\cyc\le R$. We do a similar thing for all other symmetry groups we search over.

\subsection{Cyclic + transpose symmetry}
\begin{definition}
Under $G:=\ang{\cyc,\tp}$ symmetry, we have the following orbits:
\begin{itemize}
    \item $\begin{aligned}
    \Theta_\id(A,B,C)=\big[ & (A,B,C),(B,C,A),(C,A,B), \\
    & (C^\intercal,B^\intercal,A^\intercal),
    (B^\intercal,A^\intercal,C^\intercal),\\
    & (A^\intercal,C^\intercal,B^\intercal)\big]
    \end{aligned}$
    \item $\begin{aligned}
    \Theta_\tp(S,H)=\big[ & (S,H,H^\intercal),(H,H^\intercal,S), \\
    & (H^\intercal,S,H)\big]
    \end{aligned}$
    \begin{itemize}
        \item[$\circ$] $S$ must satisfy $S=S^\intercal$
    \end{itemize}
    \item $\Theta_\cyc(D)=\bracket{(D,D,D),(D^\intercal,D^\intercal,D^\intercal)}$
    \item $\Theta_{\ang{\cyc,\tp}}(Z)=\bracket{(Z,Z,Z)}$
    \begin{itemize}
        \item[$\circ$] $Z$ must satisfy $Z=Z^\intercal$
    \end{itemize}
\end{itemize}

Our resulting decomposition is
\begin{align*}
& \paren{\listjoin_{0\le r<R_\id} \Theta_\id(A^{(r)},B^{(r)},C^{(r)})} \\
\listjoin & \paren{\listjoin_{0\le r<R_\tp} \Theta_\tp(S^{(r)},H^{(r)})} \\
\listjoin & \paren{\listjoin_{0\le r<R_\cyc} \Theta_\cyc(D^{(r)})} \\
\listjoin & \paren{\listjoin_{0\le r<R_{\ang{\cyc,\tp}}} \Theta_{\ang{\cyc,\tp}}(Z^{(r)})},
\end{align*}
with total rank $R=6R_\id+3R_\tp+2R_\cyc+R_{\ang{\cyc,\tp}}$.
\end{definition}

Since $\cyc\tp=\tp\cyc^{-1}$, $G=\ang{\cyc}\ang{\tp}\cong \mathrm{Sym}(3)$; to address matrix triplets $t$ with stabilizer $\ang{\cyc^k\tp}$, notice that we can define $t':=\cyc^k(t)$ s.t. $\tp(t')=t'$ and thus have $G(t)=G(t')=\ang{\cyc}(t')=\ang{\cyc}(\cyc^k(t))=\ang{\cyc}(t)$, so $t$ yields the $\Theta_\tp$ orbit type (up to permutation of matrix triplets).

Like in the $\ang{\cyc}$-symmetry case, we can force each orbit representative of $\Theta_\id$ and $\Theta_\cyc$ to be the strict lex minimum element in its corresponding orbit (this technique does not seem to work for $\Theta_\tp$).
Then we apply the merges $\Theta_\id(A,B,C),\Theta_\id(A,B,C')\rightarrow \Theta_\id(A,B,C+C')$ and $\Theta_\tp(S,H),\Theta_\tp(S',H)\rightarrow \Theta_\tp(S+S',H)$\footnote{
    This merging relies on the property that since $S$ and $S'$ are required to be symmetric, $S+S'$ will also be symmetric.
}.
Finally, we force orbit representatives of the same orbit type to be strictly lex increasing. 

Thus, we add the following constraints to the tensor decomposition SAT:
\begin{itemize}
    \item $(A^{(r)},B^{(r)},C^{(r)})\lexlt t$
    
    $\forall t\in\{
    (B,C,A),(C,A,B), \\(C^\intercal,B^\intercal,A^\intercal),(B^\intercal,A^\intercal,C^\intercal),(A^\intercal,C^\intercal,B^\intercal)
    \}$
    \item $(A^{(r)},B^{(r)})\lexlt (A^{(r+1)},B^{(r+1)})$
    
    \item $D^{(r)}\lexlt (D^{(r)})^\intercal$
    \item $D^{(r)}\lexlt D^{(r+1)}$
    
    \item $H^{(r)}\lexlt H^{(r+1)}$
    
    \item $Z^{(r)}\lexlt Z^{(r+1)}$
\end{itemize}

The constraints $S^{(r)}=(S^{(r)})^\intercal$ and $Z^{(r)}=(Z^{(r)})^\intercal$ are directly inlined by creating variables $s^{(r)}_{i,j}, z^{(r)}_{i,j}$ for all $i\le j$ and setting $S^{(r)}:=\M{s^{(r)}_{\min{i,j},\max{i,j}}}_{0\le i<n, 0\le j<n}$, $Z^{(r)}:=\M{z^{(r)}_{\min{i,j},\max{i,j}}}_{0\le i<n,\ 0\le j<n}$.

\subsection{Cyclic + involute-sandwich symmetry}
\begin{definition}
Under $\ang{\cyc,\sw{F,F,F}}$, with $F=\M{1&1&0\\0&1&0\\0&0&1}$, we have the following orbits:
\begin{itemize}
    \item $\begin{aligned} \Theta_\id(A,B,C)=\big[ & (A,B,C),(B,C,A),(C,A,B),\\
    & (FAF^{-1},FBF^{-1},FCF^{-1}),\\
    & (FBF^{-1},FCF^{-1},FAF^{-1}),\\
    & (FCF^{-1},FAF^{-1},FBF^{-1})\big]
    \end{aligned}$
    
    \item $\begin{aligned}
    \Theta_{\sw{F,F,F}}(X,Y,Z)=\big[ & (X,Y,Z),(Y,Z,X),\\
    & (Z,X,Y)\big]
    \end{aligned}$
    \begin{itemize}
        \item[$\circ$] $X,Y,Z$ must satisfy $FXF^{-1}=X$, $FYF^{-1}=Y$, $FZF^{-1}=Z$
    \end{itemize}
    \item $\begin{aligned}
    \Theta_\cyc(D)=\big[ & (D,D,D),\\
    & (FDF^{-1},FDF^{-1},FDF^{-1})\big]
    \end{aligned}$
    \item $\Theta_{\ang{\cyc,\sw{F,F,F}}}(U)=\bracket{(U,U,U)}$
    \begin{itemize}
        \item[$\circ$] $U$ must satisfy $FUF^{-1}=U$
    \end{itemize}
\end{itemize}

Our decomposition is
\begin{align*}
& \paren{\listjoin_{0\le r<R_\id} \Theta_\id(A^{(r)},B^{(r)},C^{(r)})} \\
\listjoin & \paren{\listjoin_{0\le r<R_{\sw{F,F,F}}} \Theta_{\sw{F,F,F}}(X^{(r)},Y^{(r)},Z^{(r)})} \\
\listjoin & \paren{\listjoin_{0\le r<R_\cyc} \Theta_\cyc(D^{(r)})} \\
\listjoin & \paren{\listjoin_{0\le r<R_{\ang{\cyc,\sw{F,F,F}}}} \Theta_{\ang{\cyc,\sw{F,F,F}}}(U^{(r)})},
\end{align*}
with total rank $R=6R_\id+3R_{\sw{F,F,F}}+2R_\cyc+R_{\ang{\cyc,\sw{F,F,F}}}$.
\end{definition}

Since $F^2=I$, $\sw{F,F,F}^2=\id$; then since $\sw{F,F,F}\cyc=\cyc\sw{F,F,F}$, $G\cong C_3\times C_2\cong C_6$, so the above orbits cover all possibilities.

We chose not to inline the equation $FMF^{-1}=M$ and instead just add it into the SAT instance, as inlining would likely require calculating a basis for the span of all possible matrices $M$ satisfying $FMF^{-1}=M$, then rewriting $M$ as a linear combination of the resulting basis matrices, which would make the resulting code significantly more complex.

As before, we can force each orbit representative to be the lex minimum element in its orbit, apply $\Theta_\id(A,B,C),\Theta_\id(A,B,C')\rightarrow \Theta_\id(A,B,C+C')$ and $\Theta_{\sw{F,F,F}}(X,Y,Z),\Theta_{\sw{F,F,F}}(X,Y,Z')\rightarrow \Theta_{\sw{F,F,F}}(X,Y,Z+Z')$\footnote{
    This merging relies on the property that $(FZF^{-1}=Z \ \land \ FZ'F^{-1}=Z') \Rightarrow F(Z+Z')F^{-1}=Z+Z'$.
}, and finally sort representatives of matching orbit types in strict lex increasing order.
Thus, we add the following constraints to the tensor decomposition SAT:

\begin{itemize}
    \item $(A^{(r)},B^{(r)},C^{(r)})\lexlt t$
    
    $\begin{aligned}
    \forall t\in\big\{ & (B^{(r)},C^{(r)},A^{(r)}),\ (C^{(r)},A^{(r)},B^{(r)}), \\
    & (FA^{(r)}F^{-1},FB^{(r)}F^{-1},FC^{(r)}F^{-1}),\\
    & (FB^{(r)}F^{-1},FC^{(r)}F^{-1},FA^{(r)}F^{-1}),\\
    & (FC^{(r)}F^{-1},FA^{(r)}F^{-1},FB^{(r)}F^{-1})\big\}
    \end{aligned}$
    
    \item $(A^{(r)},B^{(r)})\lexlt (A^{(r+1)},B^{(r+1)})$

    \item $FX^{(r)}F^{-1}=X^{(r)}$, $FY^{(r)}F^{-1}=Y^{(r)}$, $FZ^{(r)}F^{-1}=Z^{(r)}$
    
    \item $(X^{(r)},Y^{(r)},Z^{(r)})\lexlt t$
    
    $\forall t\in\brace{(Y^{(r)},Z^{(r)},X^{(r)}),(Z^{(r)},X^{(r)},Y^{(r)})}$
    
    \item $(X^{(r)},Y^{(r)})\lexlt (X^{(r+1)},Y^{(r+1)})$

    \item $D^{(r)}\lexlt FD^{(r)}F^{-1}$
    \item $D^{(r)}\lexlt D^{(r+1)}$

    \item $FU^{(r)}F^{-1}=U^{(r)}$
    \item $U^{(r)}\lexlt U^{(r+1)}$
\end{itemize}


\section{Conclusion}
Using Z3 as our Boolean SAT solver, we ruled out all $\ang{\cyc,\tp}$ or $\ang{\cyc,\sw{F,F,F}}$ -symmetric decompositions of $\mmt{3,3,3}$ with rank $\le 21$, as well as all $\ang{\cyc}$-symmetric decompositions with rank $\le 15$. More detail is given in Section \ref{results}.

We conclude with some suggestions on what to try next. For searching $\ang{\cyc}$-symmetric decompositions, one could add artificial sparsity constraints to rule out certain subsets of decompositions up to rank 21, although doing so efficiently is rather nontrivial \cite{sat-sparsity}.

One could also try proving that a stricter canonical form than the ones we considered can always be obtained from a given decomposition. We note that for $\ang{\cyc,\tp}$-symmetry, we had the additional constraints $S^{(r)}\ne H^{(r)}$ in an earlier implementation, but later found that these constraints made our program significantly slower on certain combinations of orbit counts and did improve total running time, so we removed them.

Another more general method to remove rank redundancy, which is related to requiring canonical forms, is to group matrix triplets $(A,B,C)$ by matching $A$ matrices, and within a group $[(A^{(r)},B^{(r)},C^{(r)})]_r$, force $[B^{(r)}]_r$ and $[C^{(r)}]_r$ to form linearly independent collections of matrices. This seems even more difficult to encode in SAT than sparsity constraints.

Finally, one could search over all the other symmetry groups from our previous work for which rank-$\le 21$ decompositions have not yet been ruled out\footnote{See footnote \ref{part2-table}.}; all such groups we listed are generated by a single transformation in $\degroote{3}$.

\section{Acknowledgments}
We thank Prof. Virginia Vassilevska Williams for her mentorship throughout the last two years in this and our previous reports on this topic.

\end{multicols}

\begin{table}
    \centering
    \begin{tabular}{|c|c|c|}
        \hline
        symmetry group & no decompositions of $\mmt{3,3,3}$ with rank $\le\dots$ & time (sec) \\
        \hline
        $\ang{\cyc}$ & 15 & $\numprint{346151}^*$ \\
        \hline
        $\ang{\cyc,\tp}$ & 21 & $\numprint{328787}^*$ \\
        \hline
        $\ang{\cyc,\sw{F,F,F}},\ F=\M{1&1&0\\0&1&0\\0&0&1}$ & 21 & $\numprint{16779}$ \\
        \hline
    \end{tabular}
    \caption{Ranks ruled out for decompositions of $\mmt{3,3,3}$ satisfying various symmetry restrictions. \\
    We used the Z3 solver and ran on a MacBook Pro 2.3 GHz 8-Core Intel Core i9, 16 GB 2667 MHz DDR4. \\
    $^*$Our program was paused several times when running on these symmetry groups. We estimate the total pause time for each case was no more than 2 hours.}
    \label{tab:results}
\end{table}

\end{document}